\documentclass{amsart}
\usepackage{amsmath,amsthm,amssymb,amscd}
\newtheorem{theorem}{Theorem}[section]
\newtheorem{lemma}[theorem]{Lemma}

\theoremstyle{definition}
\newtheorem{definition}[theorem]{Definition}
\newtheorem{example}[theorem]{Example}

\theoremstyle{remark}

\numberwithin{equation}{section}

\begin{document}

\title{The Continuity of Sequential Product of Sequential Quantum Effect Algebras}

\author{Qiang Lei}
\address{Qiang Lei. Department of Mathematics, Harbin Institute of Technology, Harbin 150001, China.}
\curraddr{}
\email{leiqiang@hit.edu.cn}


\author{Xiaochao Su}
\address{Xiaochao Su. Department of Mathematics, Harbin Institute of Technology, Harbin 150001, China.} \email{hitswh@163.com}

\author{Junde Wu}
\address{Junde Wu. Corresponding author: Department of Mathematics, Zhejiang University, Hangzhou 310027, China.}
\email{wjd@zju.edu.cn}

\curraddr{}
\email{}

%    General info

\keywords{Quantum effects, Sequential product, Continuity, Topology}

\begin{abstract}
In order to study quantum measurement theory, sequential product
defined by $A\circ B=A^{1/2}BA^{1/2}$ for any two quantum effects
$A,B$ is introduced. Physically motivated conditions ask the
sequential product to be continuous with respect to the strong
operator topology. In this paper, we study the continuity problems
of the sequential product $A\circ B=A^{1/2}BA^{1/2}$ with respect to
the other important topologies, as norm topology, weak operator
topology, order topology, interval topology, etc.
\end{abstract}
\maketitle

\section{Introduction}
\label{Introduction}

Effect algebra is an important model for studying the unsharp
quantum logic, it were introduced by D. J. Foulis and M. K. Bennett
in 1994, that is

\begin{definition}({\cite{FD94}}). A structure $(E; \oplus, 0, 1)$ is called an
effect algebra if 0, 1 are two distinguished elements and $\oplus$
is a partially defined binary operation on $E$ which satisfies the
following conditions for any $a, b, c\in E$:

(E1) If $a\oplus b$ is defined, then $b\oplus a$ is defined and
$a\oplus b=b\oplus a$.

(E2) If $a\oplus b$ and $(a\oplus b)\oplus c$ are defined, then
$b\oplus c$ and $a\oplus (b\oplus c)$ are defined and $(a\oplus
b)\oplus c=a\oplus (b\oplus c)$.

(E3) For each $a\in E$, there exists a unique $b\in E$ such that
$a\oplus b$ is defined and $a\oplus b=1$.

(E4) If $a\oplus 1$ is defined, then $a=0$.
\end{definition}

In an effect algebra $(E,0,1, \oplus)$, if $a\oplus b$ is defined,
we write $a\bot b$. For each $a\in (E,0,1, \oplus)$, it follows from
(E3) that there exists a unique element $b\in E$ such that $a\oplus
b=1$, we denote $b$ by $a'$. Let $a, b\in (E,0,1, \oplus)$, if there
exists a $c\in E$ such that $a\bot c$ and $a\oplus c=b$, then we say
that $a\leq b$ and define $c=b\ominus a$. Thus, each effect algebra
$(E,0,1, \oplus)$ has two partially defined binary operations
$\oplus$ and $\ominus$. Moreover, it follows from ({\cite{FD94}})
that $\leq $ is a partial order of $(E,0,1, \oplus)$ and satisfies
that for each $a\in E$, $0\leq a\leq 1$, $a\bot b$ if and only if
$a\leq b'$.

The most important and prototype of effect algebras is
($\mathcal{E}(\mathcal{H}), 0, I, \oplus)$, where $\mathcal{H}$ is a
complex Hilbert space, $\mathcal{E}(\mathcal{H})$ is the set of all
quantum effects, that is, all positive operators on $\mathcal{H}$
that are bounded above by the identity operator $I$, the partial
binary operation $\oplus$ is defined for $A,B\in
\mathcal{E}(\mathcal{H})$ iff $A+B\leq I$, in this case, $A\oplus
B=A+B$.

One can use quantum effects to represent the yes-no measurements
that may be unsharp ({\cite{FD94}}).

Let $\mathcal{D}(\mathcal{H})\subseteq \mathcal{B}(\mathcal{H})$ be
the set of density operators on $\mathcal{H}$, that is, the trace
class positive operators on $\mathcal{H}$ of unit trace, and
$\mathcal{P}(\mathcal{H})\subseteq \mathcal{B}(\mathcal{H})$ the set
of orthogonal projections on $\mathcal{H}$. For each $P\in
\mathcal{P}(\mathcal{H})$, there is associated a so-called
L$\ddot{u}$ders transformation
$\Phi_L^P:\mathcal{D}(\mathcal{H})\rightarrow
\mathcal{D}(\mathcal{H})$ such that for each $T\in
\mathcal{D}(\mathcal{H})$, $\Phi_L^P(T)=PTP$. Moreover, each quantum
effect $B\in \mathcal{E}(\mathcal{H})$ gives also a general
L$\ddot{u}$ders transformation $\Phi_L^B$ such that
$\Phi_L^B(T)=B^{\frac{1}{2}}TB^{\frac{1}{2}}$ (\cite{DB76,BP99}).

For $A,B\in \mathcal{E}(\mathcal{H})$, $A^{1/2}BA^{1/2}$ is called
the sequential product of $A$ and $B$ by Gudder and denoted by
$A\circ B$ (\cite{GS01,GA04,GS08}). The product $A\circ B$
represents the effect produced by first measuring $A$ then measuring
$B$. This product has also been generalized to an algebraic
structure called a sequential effect algebra (\cite{GS02}), that is

\begin{definition}({\cite{GS02}}). A sequential effect algebra is a system $(E; \oplus,\circ, 0, 1)$,  where $(E; \oplus, 0, 1)$ is an effect algebra and $\circ: E\times E\rightarrow E$ is a binary operation satisfying:

(SE1) The map $b\mapsto a\circ b$ is additive for every $a\in E$,
that is, if $b\oplus c$ is defined, then $a\circ b \oplus a\circ c$
is defined and $a\circ (b\oplus c)=a\circ b\oplus a\circ c$.

(SE2) $1\circ a=a$ for every $a\in E$.

(SE3) If $a\circ b=0$, then $a\circ b=b\circ a$.

(SE4) If $a\circ b=b\circ a$, then $a\circ b'=b'\circ a$ and $a\circ (b\circ c)=(a\circ b)\circ c$ for every $c\in E$.

(SE5) If $c\circ a=a\circ c$ and $c\circ b=b\circ c$, then $c\circ (a\circ b)=(a\circ b)\circ c$ and $c\circ (a\oplus b)=(a\oplus b)\circ c$.
\end{definition}

The operation $\circ$ is called sequential product. This product provides a mechanism for describing quantum interference because if $a\circ b\neq b\circ a$, then $a$ and $b$ interfere (\cite{GS02}).

Professor Gudder showed that for any two quantum effects $B$ and
$C$, the operation $\circ$ defined by $B\circ
C=B^{\frac{1}{2}}CB^{\frac{1}{2}}$ satisfies conditions (SE1)-(SE5),
and so is a sequential product of $\mathcal{E}(\mathcal{H})$. Thus,
$(\mathcal{E}(\mathcal{H}), 0, I, \oplus, \circ)$ is a sequential
effect algebra, we call it the sequential quantum effect algebra.

In 2005, Gudder presented 25 open problems in (\cite{GS05}) to
motive the study of sequential effect algebra theory, some of them
are solved in recent years
(\cite{LWH09,LWHW09,SJ09,SJWJD09,SJ10,SJW10,SJW09}). In 2015, Wang
etc. studied the entropies on sequential effect algebra
(\cite{WJM15}).

In \cite{GS08}, Gudder gave five physically motivated conditions
which fully characterize the sequential product on sequential
quantum effect algebra $(\mathcal{E}(\mathcal{H}), 0, I, \oplus,
\circ)$, one of the conditions asked that the sequential product
$B\circ C=B^{\frac{1}{2}}CB^{\frac{1}{2}}$ is jointly continuous
with respect to the strong operator topology. This showed that the
continuity of sequential product operation $\circ$ is an important
and interesting problem, although the continuity of the operation
$\oplus$ and $\ominus$ of effect algebras has been studied in
\cite{RZ02,RZ03,WJ05,LQ08,LQ09}, however, the continuity of the
sequential product operation $\circ$ of sequential effect algebras
has not been considered until now.

In this paper, we will fill the gap for the sequential quantum
effect algebra $(\mathcal{E}(\mathcal{H}), 0, I, \oplus, \circ)$,
that is, we will study the continuity of sequential product $B\circ
C=B^{\frac{1}{2}}CB^{\frac{1}{2}}$ on $\mathcal{E}(\mathcal{H})$
with respect to the norm topology, weak operator topology, order
convergence, order topology and interval topology. We will show that
$\circ$ on $\mathcal{E}(\mathcal{H})$ is jointly continuous with
respect to the norm topology, $\circ$ is continuous in the second
variable with respect to the weak operator topology, order
convergence, order topology and interval topology. We will present
examples to show that $\circ$ is not continuous in the first
variable with respect to the weak operator topology, order
convergence, order topology and interval topology.

\section{The jointly continuity of sequential product }

\begin{definition}.
Let $\mathcal{H}$ be a complex Hilbert Space. For any $x\in
\mathcal{H}$, the equation $P_x(T)=\|Tx\|$ defines a semi-norm $P_x$
on $\mathcal{B}(\mathcal{H})$. The family of all semi-norms
$\{P_x:x\in \mathcal{H}\}$ gives rise to a topology on
$\mathcal{B}(\mathcal{H})$ called strong operator topology and
denoted by $SOT$.
\end{definition}

In the strong operator topology, an element $T_0\in
\mathcal{B}(\mathcal{H})$ has a base of neighborhoods consisting of all sets of type
$$V(T_0:x_1,\cdots,x_m;\varepsilon)=\{T\in
\mathcal{B}(\mathcal{H}):\|(T-T_0)x_j\|<\varepsilon,
j=1,\cdots,m\},$$ where $\varepsilon$ is a positive number and
$x_1,\cdots,x_m\in \mathcal{H}$.

It can be proved
$T_\alpha\xrightarrow{SOT}T\Leftrightarrow \forall x\in
\mathcal{H}$, ~$\|(T_\alpha-T)x\| \rightarrow  0$.

Gudder had pointed out that $\circ$ is jointly continuous in the
strong operator topology(\cite{GS08}).

Next, we prove $\circ$ is continuous with respect to the norm
topology.

\begin{lemma}\label{l:202}({\cite{KRV91}}).
Let $\{A_\alpha\}_{\alpha\in \Lambda}$ be a net in $\mathcal{B}(\mathcal{H})$ and $A\in \mathcal{B}(\mathcal{H})$, $A_\alpha\geq 0,A\geq 0$.

(1) If $\|A_\alpha-A\|\rightarrow 0$, then $\|A_\alpha^{1/2}-A^{1/2}\|\rightarrow 0$.

(2) If $A_\alpha\xrightarrow{SOT} A$, then $A_\alpha^{1/2}\xrightarrow{SOT}A^{1/2}$.
\end{lemma}

\begin{theorem}
The sequential product $B\circ C=B^{\frac{1}{2}}CB^{\frac{1}{2}}$ on
sequential quantum effect algebra $(\mathcal{E}(\mathcal{H}), 0, I,
\oplus, \circ)$ is jointly continuous with respect to the norm
topology. That is, if $A_\alpha\xrightarrow{\|\cdot\|}A$ and
$B_\alpha\xrightarrow{\|\cdot\|}B$, then $A_\alpha\circ
B_\alpha\xrightarrow{\|\cdot\|}A\circ B$.
\end{theorem}

$(\mathcal{E}(\mathcal{H}), 0, I, \oplus, \circ)$, one of the
conditions asked that the sequential product $B\circ
C=B^{\frac{1}{2}}CB^{\frac{1}{2}}$

\begin{proof}
By Lemma \ref{l:202}, we have $A_\alpha^{1/2}\xrightarrow{\|\cdot\|} A^{1/2}$. Then

\begin{eqnarray*}
&&\|A_\alpha\circ B_\alpha-A\circ B\|=\|A_\alpha^{1/2}B_\alpha A_\alpha^{1/2}-A^{1/2}B A^{1/2}\|\\
&\leq &\|A_\alpha^{1/2}B_\alpha A_\alpha^{1/2}-A_\alpha^{1/2}B_\alpha A^{1/2}+A_\alpha^{1/2}B_\alpha A^{1/2}-A_\alpha^{1/2}BA^{1/2}+A_\alpha^{1/2}BA^{1/2}-A^{1/2}B A^{1/2}\|\\
&\leq& \|A_\alpha^{1/2}B_\alpha\| \|A_\alpha^{1/2}-A^{1/2}\|+\|A_\alpha^{1/2}\| \|B_\alpha-B\| \|A^{1/2}\|+\|A_\alpha^{1/2}-A^{1/2}\| \|BA^{1/2}\|.
\end{eqnarray*}
As $\|A_\alpha^{1/2}B_\alpha\|\leq 1$, $\|A_\alpha^{1/2}\| \|A^{1/2}\|\leq 1$ and $ \|BA^{1/2}\|\leq 1$, we have

$$\|A_\alpha\circ B_\alpha-A\circ B\|
\leq \|A_\alpha^{1/2}-A^{1/2}\|+ \|B_\alpha-B\| +\|A_\alpha^{1/2}-A^{1/2}\|\rightarrow 0.$$

That is $A_\alpha\circ B_\alpha\xrightarrow{\|\cdot\|}A\circ B$.
\end{proof}

\section{The continuity of the sequential product in the second variable}
\begin{definition}(\cite{KRV91}).
Suppose that $\mathcal{V}$ is a linear space with scalar field $K$, and $\mathcal{F}$ is a family of linear functionals on $\mathcal{V}$, which separates the points of $\mathcal{V}$. For any $\rho\in \mathcal{F}$, the equation $P_\rho(x)=|\rho(x)|$ defines a semi-norm $P_\rho$ on $\mathcal{V}$. The topology generated by $\{P_\rho|\rho\in
\mathcal{F}\}$ is called weak topology induced by $\mathcal{F}$.
\end{definition}

\begin{definition}(\cite{KRV91}).
The weak operator topology on $\mathcal{B}(\mathcal{H})$ is the weak
topology induced by the family $\mathcal{F}_w$ of linear functionals
$\omega_{x,y}: \mathcal{B}(\mathcal{H})\rightarrow \mathbb{C}$
defined by the equation
 ~$\omega_{x,y}(A)=\langle
Ax,y\rangle $, $x,\ y\in \mathcal{H}$. The weak operator topology is denoted by $WOT$.
\end{definition}

The family of sets of the form
$$V(T_0:\omega_{x_1,y_1},\cdots,\omega_{x_m,y_m};\varepsilon)=\{T\in
\mathcal{B}(\mathcal{H}):|\langle(T-T_0)x_j,
y_j\rangle|<\varepsilon, j=1,\cdots,m\},$$ where $\varepsilon$ is
positive number and $x_1,\cdots,x_m, y_1,\cdots,y_m\in \mathcal{H}$
constitutes a base of neighborhoods of $T_0$ in WOT.

It can be proved that $T_\alpha\xrightarrow{WOT}T\Leftrightarrow
\forall x,y\in \mathcal{H}$, $\langle T_\alpha x,y\rangle
\rightarrow \langle T x,y\rangle \Leftrightarrow \forall x\in
\mathcal{H}$, $\langle T_\alpha x,x\rangle \rightarrow \langle T
x,x\rangle$.

\begin{theorem}
The sequential product $B\circ C=B^{\frac{1}{2}}CB^{\frac{1}{2}}$ on
sequential quantum effect algebra $(\mathcal{E}(\mathcal{H}), 0, I,
\oplus, \circ)$ is continuous in the second variable with respect to
the weak operator topology. That is, if
$B_\alpha\xrightarrow{WOT}B$, then $A\circ
B_\alpha\xrightarrow{WOT}A\circ B$ for each $A\in
\mathcal{E}(\mathcal{H})$.
\end{theorem}

\begin{proof}
As $B_\alpha\xrightarrow{WOT}B$, $\langle B_\alpha x,x\rangle\rightarrow \langle Bx,x\rangle$ for each $x\in \mathcal{H}$. Then
$\langle A\circ B_\alpha x,x\rangle=\langle A^{1/2}B_\alpha A^{1/2}x,x\rangle=\langle B_\alpha A^{1/2}x, A^{1/2}x\rangle\rightarrow \langle BA^{1/2}x, A^{1/2}x\rangle=\langle A^{1/2}BA^{1/2}x,x\rangle=\langle A\circ B x,x\rangle$ for each $x\in\mathcal{H}$.
That is $A\circ B_\alpha\xrightarrow{WOT}A\circ B$.
\end{proof}

We give an example to show that the continuity of $B\circ
C=B^{\frac{1}{2}}CB^{\frac{1}{2}}$ is not correct in the first
variable with respect to WOT.

\begin{example}\label{e:301}
Let $\mathcal{H}$ be the complex separable Hilbert space $l^2$ and
$\{e_n\}_{n=1}^\infty$ be its orthonormal  basis. For each $n$,
define

\[
P_ne_i = \begin{cases}
\frac{1}{2}e_1+\frac{1}{2}e_{n+1},&i=1,\ i=n+1,\\
0,&  others.
\end{cases}
\]
and
\[
P_0e_i = \begin{cases}
\frac{1}{2}e_1,&i=1,\\
0,&  others.
\end{cases}
\]
It is easy to show that $P_n$ is an orthogonal projection operator
for each $n$. That is $P_n\xrightarrow{WOT}P_0$ is clear.

Let
\[
Be_i = \begin{cases}
\frac{1}{2}e_1+\frac{1}{2}e_2,&i=1,\ i=2,\\
0,&  others.
\end{cases}
\]
Then $B\in \mathcal{E}(\mathcal{H})$.
Since $\{P_n\}$ are orthogonal projection operators,

$$\langle P_n\circ B x,x\rangle=\langle P_n^{{\frac{1}{2}}}BP_n^{\frac{1}{2}}x,x \rangle=\langle BP_nx,P_nx\rangle\rightarrow\langle \frac{1}{4} P_0x,x\rangle$$ for each $x\in l^2$. That is $P_n\circ B\xrightarrow{WOT}\frac{1}{4}P_0$. However, $P_0\circ B=\frac{1}{2}P_0$. So
$P_n\circ B$ is not convergent to $P_0\circ B$ with respect to WOT.
\end{example}

Let $(P,\leq)$ be a poset. If $\{a_\alpha\}_{\alpha\in \Lambda}$ is
a net of $P$ and $a_\alpha\leq a_\beta$ when
$\alpha,\beta\in\Lambda$ and $\alpha\preceq \beta$, then we write
$a_\alpha \uparrow $. Moreover, if $a$ is the supremum of
$\{a_\alpha\}_{\alpha\in \Lambda}$, i.e. $a =
\vee\{a_\alpha:\alpha\in \Lambda\}$, then we write $a_\alpha
\uparrow a$. Similarly, we may write $a_\alpha \downarrow$ and
$a_\alpha \downarrow a$.

We say that a net $\{a_\alpha\}_{\alpha\in \Lambda}$ of $P$ is order convergent to $a\in P$ if there exist two nets $\{u_\alpha\}_{\alpha\in \Lambda}$ and $\{v_\alpha\}_{\alpha\in \Lambda}$ of $P$ such that $a\uparrow u_\alpha\leq a_\alpha\leq v_\alpha \downarrow a$. We denote order convergence as $a_\alpha \xrightarrow{o} a$. It can be proved that $a_\alpha \xrightarrow{o} a\Rightarrow a_\alpha \xrightarrow{SOT} a$ (\cite{MZH09}).

\begin{lemma} (\cite{KRV91}\label{l:301}). If $\{A_\alpha\}$ is a monotone increasing net of self-adjoint operators on a Hilbert space $\mathcal{H}$ and $A_\alpha\leq I$ for all $\alpha$, then $\{A_\alpha\}$ is strong-operator convergent to a self-adjoint operator $A$, and $A$ is the least upper bound of $\{A_\alpha\}$.
\end{lemma}

\begin{theorem}
The sequential product $B\circ C=B^{\frac{1}{2}}CB^{\frac{1}{2}}$ on
sequential quantum effect algebra $(\mathcal{E}(\mathcal{H}), 0, I,
\oplus, \circ)$ is continuous in the second variable with respect to
the order convergence. That is, if $B_\alpha\xrightarrow{o}B$, then
$A\circ B_\alpha\xrightarrow{o}A\circ B$.
\end{theorem}

\begin{proof}
Let $B_\alpha\xrightarrow{o}B$. Then there exist two nets
$\{C_\alpha\},\{D_\alpha\}$ such that $C_\alpha\uparrow B$ and
$D_\alpha\downarrow B$ satisfying $C_\alpha\leq B_\alpha\leq
D_\alpha$. It follows that $A^{\frac{1}{2}}C_\alpha
A^{\frac{1}{2}}\leq A^{\frac{1}{2}}B_\alpha A^{\frac{1}{2}}\leq
A^{\frac{1}{2}}D_\alpha A^{\frac{1}{2}}$. That is $A\circ
C_\alpha\leq A\circ B_\alpha\leq A\circ D_\alpha$. It is clear that
$A\circ C_\alpha \uparrow$ and $A\circ D_\alpha \downarrow$. Since
the order convergence is stronger than SOT, we have
$C_\alpha\xrightarrow{SOT}B$ and $D_\alpha\xrightarrow{SOT}B$. From
the fact that $\circ$ is jointly continuous with respect to $SOT$,
it follows that $A\circ C_\alpha\xrightarrow{SOT} A\circ B$ and
$A\circ D_\alpha\xrightarrow{SOT} A\circ B$. By Lemma \ref{l:301},
$A\circ C_\alpha \uparrow A\circ B$ and $A\circ D_\alpha \downarrow
A\circ B$. That is, $$A\circ B\uparrow A\circ C_\alpha\leq A\circ
B_\alpha\leq A\circ D_\alpha\downarrow A\circ B.$$ Therefore,
$A\circ B_\alpha\xrightarrow{o} A\circ B$.
\end{proof}

However, the conclusion is not correct in the first variable. That
is,

\begin{example}\label{e:302}
Let $A_n=I-\frac{1}{n}
\left(\begin{array}{cc}
1 & 1 \\
1 & 1
\end{array}\right)
$ and $B=\left(\begin{array}{cc}
1 & 0 \\
0 & 0
\end{array}\right)$. Then $A_n\uparrow I$ and

$A_n^{1/2}=\frac{1}{2}\left(\begin{array}{cc}
\sqrt{1-\frac{2}{n}}+1 & \sqrt{1-\frac{2}{n}}-1 \\
\sqrt{1-\frac{2}{n}}-1 & \sqrt{1-\frac{2}{n}}+1
\end{array}\right)$,

$A_n\circ B=A_n^{1/2}BA_n^{1/2}=\frac{1}{2}\left(\begin{array}{cc}
1-\frac{1}{n}+\sqrt{1-\frac{2}{n}} & -\frac{1}{n} \\
-\frac{1}{n} & 1-\frac{1}{n}-\sqrt{1-\frac{2}{n}}
\end{array}\right)$,

$\langle A_n\circ Bx,x\rangle=\frac{1}{2}[(1-\frac{1}{n}+\sqrt{1-\frac{2}{n}})x_1^2+(1-\frac{1}{n}-\sqrt{1-\frac{2}{n}})x_2^2-\frac{2}{n}x_1x_2]$,

$\langle I\circ Bx,x\rangle=\langle Bx,x\rangle=x_1^2$.

Suppose $A_n\circ B\xrightarrow{o} I\circ B=B$. Then there exists an increasing net $\{C_n\}\subseteq \mathcal{E}(\mathcal{H})$ and a decreasing net $\{D_n\}\subseteq\mathcal{E}(\mathcal{H})$ satisfying $B\uparrow C_n\leq A_n\circ B\leq D_n \downarrow B$.

Let $C_n=\left(\begin{array}{cc}
a_n & b_n \\
b_n & c_n
\end{array}\right)$. Then $\langle C_nx,x\rangle\leq \langle Bx,x\rangle$ for each $x$. It follows that $b_n=c_n=0$, $a_n\uparrow 1$ and $C_n=\left(\begin{array}{cc}
a_n & 0 \\
0 & 0
\end{array}\right)$ where $a_n\geq 0$ and $a_n\uparrow 1$. $\langle C_nx,x\rangle=a_nx_1^2$. For each $x=\left(\begin{array}{c}
x_1 \\
x_2
\end{array}\right)$ with $x_1\neq 0$,
\begin{eqnarray*}
 & & \langle (C_n-A_n\circ B)x,x\rangle\\
&=& [a_n-\frac{1}{2}(1-\frac{1}{n}+\sqrt{1-\frac{2}{n}})]x_1^2-
\frac{1}{2}(1-\frac{1}{n}-\sqrt{1-\frac{2}{n}})x_2^2+\frac{1}{n}x_1x_2] \\
    &=& \frac{1}{2}x_1^2[-(1-\frac{1}{n}-\sqrt{1-\frac{2}{n}})(\frac{x_2}{x_1})^2+
\frac{2}{n}(\frac{x_2}{x_1})+2a_n-(1-\frac{1}{n}+\sqrt{1-\frac{2}{n}})].
\end{eqnarray*}
Let $t=\frac{x_2}{x_1}$. Consider the function
$$f(t)=-(1-\frac{1}{n}-\sqrt{1-\frac{2}{n}})t^2+\frac{2}{n}t+2a_n-(1-\frac{1}{n}+
\sqrt{1-\frac{2}{n}}).$$ $\Delta=(\frac{2}{n})^2+4(1-\frac{1}{n}-
\sqrt{1-\frac{2}{n}})[2a_n-(1-\frac{1}{n}+
\sqrt{1-\frac{2}{n}})]=8a_n(1-\frac{1}{n}- \sqrt{1-\frac{2}{n}})>
0$. So there exists a $t$ such that $f(t)>0$. Therefore, there
exists an $x$ such that $\langle (C_n-A_n\circ B)x,x\rangle>0$. This
contradicts $C_n\leq A_n\circ B$. Thus, we have $\{A_n\circ B\}$ is
not order convergence to $I\circ B=B$.
\end{example}

Let $(P,\leq)$ be a poset. Denote $\mathcal{F}=\{F\subseteq P:\ if \ \{a_\alpha\}_{\alpha\in \Lambda}\subseteq F$ is a net and $\{a_\alpha\}_{\alpha\in \Lambda}$ is order convergent to $a\in P$, then $a\in F\}$. It can be proved that the family $\mathcal{F}$ of subsets of $P$ defines a topology $\tau_o$ on $P$ such that $\mathcal{F}$ consists of all closed sets of this topology.
The topology $\tau_o$ is called the order topology on $P$ (\cite{RZ03}).

It can be proved that the order topology $\tau_o$ of $P$ is the finest topology on $P$ such that for each net $\{a_\alpha\}_{\alpha\in \Lambda}$ of $P$, if $a_\alpha\xrightarrow{o} a$, then $a_\alpha\xrightarrow{\tau_o} a$. But the converse is not necessarily true (\cite{RZ03}).

\begin{theorem}
The sequential product $B\circ C=B^{\frac{1}{2}}CB^{\frac{1}{2}}$ on
sequential quantum effect algebra $(\mathcal{E}(\mathcal{H}), 0, I,
\oplus, \circ)$ is continuous in the second variable with respect to
the order topology. That is, if $B_\alpha\xrightarrow{\tau_o}B$,
then $A\circ B_\alpha\xrightarrow{\tau_o}A\circ B$ for each $A\in
\mathcal{E}(\mathcal{H})$.
\end{theorem}

\begin{proof} Firstly, let $f:\mathcal{E}(\mathcal{H})\rightarrow \mathcal{E}(\mathcal{H})$
defined by $f(B)=A\circ B=A^{1/2}BA^{1/2}$, $F$ be a closed set with
respect to the order topology $\tau_o$, $F_1=f^{-1}(F)=\{B\in
\mathcal{E}(\mathcal{H}): A^{1/2}BA^{1/2}\in F\}$. Next, we prove
that $F_1$ is a closed set with respect to the order topology
$\tau_o$. Let $\{B_\alpha\}\subseteq F_1$ and
$B_\alpha\xrightarrow{o} B$. Then $A^{1/2}B_\alpha
A^{1/2}\xrightarrow{o}A^{1/2}B A^{1/2}$ since $\circ$ is continuous
in the second variable with respect to the order convergence. Note
that order convergence is stronger than order topology, we have
$A^{1/2}B_\alpha A^{1/2}\xrightarrow{\tau_o}A^{1/2}B A^{1/2}$. As
$\{A^{1/2}B_\alpha A^{1/2}\}\subseteq F$ and $F$ is closed in
$\tau_o$, we obtain $A^{1/2}BA^{1/2}\in F$. Thus $B\in F_1$ and
$F_1$ is closed in $\tau_o$. Therefore $f$ is continuous according
to $\tau_o$. That is $B_\alpha\xrightarrow{\tau_o} B$ implies that
$A\circ B_\alpha\xrightarrow{\tau_o}A\circ B$ for each $A\in
\mathcal{E}(\mathcal{H})$.
\end{proof}

Now, we show also that the conclusion is not correct in the first
variable.

\begin{example}
Let $\{A_n\}$ and $B$ be defined as the same in Example \ref{e:302}. Then $A_n\uparrow I$ implies $A_n\xrightarrow{\tau_o}I$. Suppose $f(A)=A\circ B$ and $f$ is continuous with respect to $\tau_o$. It follows that $A_n\circ B\xrightarrow{\tau_o}I\circ B=B$. Denote $F=\{A_n\circ B\}$. If $\{A_n\circ B\}$ is order convergent and $A_n\circ B\xrightarrow{o}M$, then $\langle A_n\circ Bx,x\rangle\rightarrow \langle Mx,x\rangle$ for each $x$ since the order convergence is stronger than WOT. As in Example \ref{e:302}, $\langle A_n\circ Bx,x\rangle\rightarrow \langle Bx,x\rangle$. It follows that $M=B$ which is contradict with Example \ref{e:302}. Thus $\{A_n\circ B\}$ is not order convergent and $F=\{A_n\circ B\}$ is closed in $\tau_o$ by the definition. Let $F_1=f^{-1}(F)=\{A\in\mathcal{E}(\mathcal{H}): A\circ B\in F\}$. Then $F_1$ is closed with respect to $\tau_o$ as we have supposed $f$ is continuous. As $\{A_n\}\subseteq F_1$ and $A_n\xrightarrow{o}I$, we have $I\in F_1$. This implies $B\in F$. This is a contradiction. So $f$ is not continuous with respect to $\tau_o$.
\end{example}

By the interval topology of a poset $P$, we mean the topology which
is defined by taking all closed intervals [a, b] as a sub-basis of closed sets of $P$.
We denote by $\tau_I$ the interval topology. It can be verified that
each closed interval [a, b] of a poset $P$ is a closed set with respect to the
order topology of $P$, so the interval topology is weaker than the
order topology (\cite{LQ09}).

\begin{lemma} ({\cite{LQ09}}\label{l:302}).
Let $(P,\leq)$ be a poset and $\{a_\alpha\}_{\alpha\in \Lambda}$ be
a net in $(P,\leq)$. Then $a_\alpha\xrightarrow{\tau_I} a$ iff for
any subnet $\{a_\gamma\}_{\gamma\in\Upsilon}$, $a_\gamma\geq r$ for
$r\in P$ implies $a\geq r$ and $a_\gamma\leq r$ for $r\in P$ implies
$a\leq r$.
\end{lemma}

\begin{theorem}
The sequential product $B\circ C=B^{\frac{1}{2}}CB^{\frac{1}{2}}$ on
sequential quantum effect algebra $(\mathcal{E}(\mathcal{H}), 0, I,
\oplus, \circ)$ is continuous in the second variable with respect to
the order topology. That is, if $B_\alpha\xrightarrow{\tau_I}B$,
then $A\circ B_\alpha\xrightarrow{\tau_I}A\circ B$ for each $A\in
\mathcal{E}(\mathcal{H})$.
\end{theorem}

\begin{proof}
Let $\{B_\gamma\}$ be any subnet of $\{B_\alpha\}$ and $A\circ
B_\gamma\geq C_1$ for $A,C_1\in \mathcal{E}(\mathcal{H})$. That is
$A^{1/2}B_\gamma A^{1/2}\geq C_1$. For any $\lambda>0$, $(\lambda
I+A)^{1/2}B_\gamma (\lambda I+A)^{1/2}\geq C_1$ and $(\lambda
I+A)^{1/2}$ is invertible. Then we obtain $$B_\gamma\geq(\lambda
I+A)^{-1/2}C_1(\lambda I+A)^{-1/2}$$ for each $\gamma$. As
$B_\alpha\xrightarrow{\tau_I} B$, by Lemma \ref{l:302}, we have
$$B\geq (\lambda I+A)^{-1/2}C_1(\lambda I+A)^{-1/2}.$$ So $$(\lambda
I+A)^{1/2}B (\lambda I+A)^{1/2}\geq C_1.$$ Let $\lambda\rightarrow
0$, we obtain $A^{1/2}BA^{1/2}\geq C_1$. That is $A\circ B\geq C_1$.

Next, let $A\circ B_\gamma\leq C_2$. Namely, $A^{1/2}B_\gamma A^{1/2}\leq C_2$. Let $\lambda>0$. It is easy to prove that $(\lambda I+A)^{1/2}\leq \sqrt{\lambda}I+A^{1/2}$. So
\begin{eqnarray*}
  (\lambda I+A)^{1/2}B_\gamma (\lambda I+A)^{1/2} &\leq& (\sqrt{\lambda}I+A^{1/2})B_\gamma (\sqrt{\lambda}I+A^{1/2})\\
&=& \lambda B_\gamma+\sqrt{\lambda}(A^{1/2}B_\gamma+B_\gamma A^{1/2})+A^{1/2}B_\gamma A^{1/2}
\end{eqnarray*}
It is also easy to prove $\sqrt{\lambda}(A^{1/2}B_\gamma+B_\gamma
A^{1/2})\leq 2\sqrt{\lambda}I$. So $$(\lambda I+A)^{1/2}B_\gamma
(\lambda I+A)^{1/2}\leq (\lambda+2\sqrt{\lambda})I+C_2.$$ Since
$(\lambda I+A)^{1/2}$ is invertible, it follows $$B_\gamma\leq
(\lambda I+A)^{-1/2}[(\lambda+2\sqrt{\lambda})I+C_2](\lambda
I+A)^{-1/2}.$$ As $B_\alpha\xrightarrow{\tau_I} B$, $$B\leq (\lambda
I+A)^{-1/2}[(\lambda+2\sqrt{\lambda})I+C_2](\lambda I+A)^{-1/2}$$
and $$(\lambda I+A)^{1/2}B(\lambda I+A)^{1/2}\leq
(\lambda+2\sqrt{\lambda})I+C_2.$$ Let $\lambda\rightarrow 0$, we
have $A^{1/2}BA^{1/2}\leq C_2$. That is $A\circ B\leq C_2$. From
Lemma \ref{l:302} we obtain $A\circ
B_\alpha\xrightarrow{\tau_I}A\circ B$.
\end{proof}

However, the conclusion is not correct in the first variable, too.

\begin{lemma}(\cite{KRV91}\label{l:303}).
The set $\mathcal{P}(\mathcal{H})$ of orthogonal projections on
$\mathcal{H}$ is weak-operator dense in the set
$\mathcal{B}(\mathcal{H})_1^{+}$ of positive operators in the unit
ball of $\mathcal{B}(\mathcal{H})$.
\end{lemma}

\begin{example}
For $\frac{I}{2}$,by Lemma \ref{l:303}, there exists a sequence of
projections $\{E_n\}$ such that $E_n\xrightarrow{WOT} \frac{I}{2}$.
As $WOT$ is stronger than $\tau_I$, it follows that
$E_n\xrightarrow{\tau_I} \frac{I}{2}$. For some $x_0$ with
$\|x_0\|=1$, denote $\mathcal{V}=\{F\in \mathcal{B}(\mathcal{H}):
|\langle (\frac{I}{2}-F)x_0,x_0\rangle| <\frac{1}{3}\}$. Then $V$ is
a neighborhood of $\frac{I}{2}$ for WOT. Without lost generality, we
suppose that $E_n\in \mathcal{V}$ for each $n$. It follows that
$$\langle\frac{I}{2}x_0,x_0\rangle-\langle E_nx_0,x_0\rangle\leq
|\langle(\frac{I}{2}-E_n)x_0,x_0\rangle|<\frac{1}{3}.$$ That is
$\langle E_nx_0,x_0\rangle\geq\frac{1}{2}-\frac{1}{3}=\frac{1}{6}$.
So $\bigwedge E_n\neq 0$. Denote $B=\bigwedge E_n$, then $B$ is an
orthogonal projection. $E_n\circ B=E_nBE_n=B$, $\frac{I}{2}\circ
B=\frac{1}{2}B$. $E_n\circ B=B\geq B$. However,$\frac{I}{2}\circ
B=\frac{1}{2}B$. By Lemma \ref{l:302}, $\{E_n\circ B\}$ is not
convergent to $\frac{I}{2}\circ B$ with respect to $\tau_I$.
\end{example}

\thanks{{\bf Acknowledgement}: This  project is supported by National Natural Science Foundation of China (11101108, 11171301, 11571307)
and by the Doctoral Programs Foundation of the Ministry of Education of China (J20130061).}

\bibliographystyle{amsplain}

\end{document}